\documentclass[prb,aps,twocolumn,showpacs]{revtex4}
\usepackage{epsfig,latexsym,amssymb,amsmath,amsbsy,graphics,graphicx}

\begin{document}

\title{Random Antiferromagnetic Spin-1/2 Chains with Competing Interactions}

\author{Eddy Yusuf and Kun Yang}
\affiliation{
National High Magnetic Field Laboratory and Department of Physics,
Florida State University, Tallahassee, Florida 32306
}
\date{\today}
\begin{abstract}
We study disordered antiferromagnetic spin-1/2 chains with nearest- and 
further-neighbor
interactions using the real-space renormalization-group method. We find
that the system supports two different phases, depending on the ratio of the 
strength between nearest-neighbor and further-neighbor interactions as well
the bond randomness strength. For weak further neighbor coupling the system is
in the familiar random singlet phase, while stronger further neighbor coupling
drives the system to a
large spin phase similar to that found in the study of random
antiferromagnetic-ferromagnetic
spin chains. The appearance of the
large spin phase in the absence of ferromagnetic coupling
is due to the frustration introduced by further neighboring 
couplings, and is unique to the disordered chains. 
\end{abstract}
\maketitle

\section{Introduction}

One-dimensional (1D) quantum spin systems have been of interest to physicists
for many years. This is not only because these systems have been good 
testing grounds for various theoretical techniques and approximations but also
because they exhibit a wealth of fascinating low-energy physics.
Among various intriguing phenomena of these systems, 
the interplay between quantum 
fluctuation and disorder has attracted considerable recent attention. The most
thoroughly studied model in this context is the random
antiferromagnetic (AF) spin-1/2 chain with 
nearest neighbor interaction. It has been shown,\cite{fisher} using the
celebrated real space renormalization group (RSRG) method,\cite{mdh,bhatt}
that the low-energy physics of the model is controlled by the random singlet 
(RS)
fixed point of the RSRG and is universal. Among the universal properties of the
random singlet phase are the uniform spin susceptibility: 
$\chi\sim 1/T\log^2T$,
and the disorder averaged spin-spin correlation function: 
$\langle{\bf S}_i\cdot{\bf S}_j\rangle\sim (-1)^{i-j}/(i-j)^2$.
The RSRG method (with proper extensions) has also been
applied with considerable success 
to a number of other disordered spin chain 
models (all with nearest neighbor interaction only),
\cite{fisher2,westerberg,hybg,boechat,hy,monthus,yb,damle,refael,saguia} 
as well as two-leg spin ladders.\cite{melin,ek,ek2}

In the present work we study random AF spin-1/2 chains with nearest {\em and}
further neighbor couplings, using the RSRG method. Our motivation comes from
the following considerations. First of all, as mentioned above, existing 
theoretical studies 
have been focusing on models with nearest neighbor couplings
only; the RG flow equations
of the couplings are relatively simple in this case which allows, for example,
exact analytical solution of the fixed point in the case of random AF spin-1/2
chain.\cite{fisher} In real physical systems, on the other hand, further 
neighbor couplings are always present, and in certain cases they can even be
quite strong. There are a few promising experimental realizations of materials
that exhibit non trivial next-nearest neighbor interactions. 
One of the examples of real physical systems that may meet the criteria is 
CuGeO$_3$.\cite{lorenzo,riera,chakravarty,regnault,mizuno} Studies on 
this system have revealed that the angle of Cu-O-Cu bond is close to 
90$^{\circ}$. This will induce a competition between antiferromagnetic 
superexchange between the Cu ions mediated by the oxygen ion and ferromagnetic 
direct exchange between the Cu ions. As a result the nearest-neighbor 
superexchange interaction is weakened and hence it is expected that the 
next-nearest-neighbor interactions which arise from the Cu-O-O-Cu path 
cannot be neglected. The strength of the second-neighbor bonds can also be 
controlled by applying pressure to such systems. Masuda and 
coworker\cite{masuda} 
studied the effect of pressure on highly Mg-doped CuGeO$_3$ and
found that the frustration is enhanced as the pressure is increased. Another
example of material that exhibits non trivial second-neighbor interaction is 
Cu$_6$Ge$_6$ O$_{18}-x$H$_2$O studied by Hase and coworker.\cite{hase} 
Motivated by these experimental realizations, we would thus like to study 
the effects of next-nearest-neighbor interactions, and in particular, 
the stability of the RS fixed point against their presence.

Secondly, nearest neighbor models have no frustration in them. Further
neighbor interactions, on the other hand, can introduce frustration, and this
is known to lead to new physics and phases in the case of pure chains.  
For example, it is known in the case of spin-1/2 chain with nearest and next
nearest neighbor couplings ($J_1$ and $J_2$), 
there are two different phases depending on the
ratio between the two.\cite{haldane, kuboki, tonegawa, affleck, okamoto}
For zero or small $J_2/J_1$, the 
system is in a gapless (critical) phase with power-law spin-spin correlation, 
while for larger $J_2/J_1$ the system spontaneously dimerizes and opens a gap
in the excitation spectrum, and the spin-spin correlation becomes 
short-ranged. In the special case of $J_2/J_1=1/2$, which is the so-called
Majumdar-Ghosh model, the ground state of the system
is known exactly; they are collections of neighboring spins forming singlet 
pairs over either even or
odd nearest neighbor bonds.\cite{majumdar, mg, broek, ss} 
It is thus of interest to study how frustration affects the physics of 
disordered chains, and whether new phases can be stabilized by it.

Our results can be summarized as follows. We find that there are two phases
in the model we are considering, controlled by the ratio of the strength of 
nearest-neighbor and next-nearest neighbor interactions and the strength of
bond randomness. The Random Singlet (RS) phase is found to be stable 
against weak further neighbor couplings; in this case the strength of further 
neighbor couplings (as measured by strength of nearest neighbor couplings)
flow to zero as energy scale decreases, thus the low-temperature properties of
the system is
still controlled by the Random Singlet (RS) fixed point. 
For 
strong enough further neighbor couplings, on the other hand, the RS phase
becomes unstable and the system is 
driven into 
another phase which is controlled by large effective spins at low energies. 
We find that in this phase the system is still dominated by effective 
nearest-neighbor interactions at low-energy; however the effective couplings
can be either antiferromagnetic or 
ferromagnetic, with random distributions. We conclude that this phase is the 
same as that found in random antiferromagnetic (AF)-ferromagnetic (F)
spin chain systems with 
nearest-neighbor interactions only studied by 
Westerberg {\em et al.}.\cite{westerberg}
The physical origin of the appearance of effective ferromagnetic couplings is
the frustration introduced by further neighbor couplings.

The remainder of the paper is organized as follows. In Sec. II we introduce 
the model we study and discuss the application of RSRG to this model. Results 
of our numerical studies on the model are presented in Sec. III. 
In Sec. IV we 
summarize our findings and make contact with previous works that are related 
to our studies.

\section{The Model}
\begin{figure*}
\includegraphics[scale=0.4]{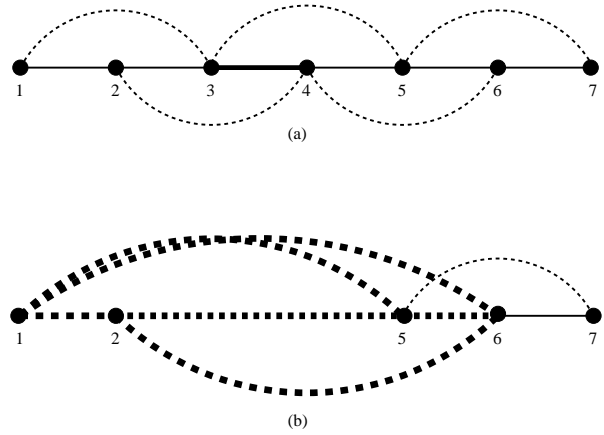}
\caption{(a) Schematic diagram for the AF spin-1/2 chain given by the 
Hamiltonian (\ref{hamiltonian}). In addition to the nearest-neighbor 
couplings between the spins, we also include the next-nearest-neighbor 
couplings represented by the dashed lines. Here the strongest bond is 
represented by the thick bold line. (b) The renormalization scheme after
the strongest bond is decimated. The thick dashed lines are the renormalized
couplings.}
\label{fig:chain}
\end{figure*}

We consider the antiferromagnetic (AF) spin-1/2 chain described by the 
following Hamiltonian :
\begin{equation}
H = \sum_{i=1}^{N-1}J_i \boldsymbol{S}_i \cdot \boldsymbol{S}_{i+1} + 
\sum_{i=1}^{N-2}K_i \boldsymbol{S}_i \cdot \boldsymbol{S}_{i+2},
\label{hamiltonian}
\end{equation}
where $N$ is the number of spins on the chain, $\boldsymbol{S}_i$ is a 
spin-1/2 operator at the $i^{th}$ site and the positive couplings $J_i$ and 
$K_i$ are distributed randomly according to some probability distributions 
which will be described in more details in the next section.
The Hamiltonian written down in Eq. (\ref{hamiltonian}) consists of two terms,
where the first term describes nearest-neighbor interactions between the spins
and the second term describes next-nearest-neighbor 
(n.n.n.) interactions. The schematic diagram of the system described by the 
Hamiltonian (\ref{hamiltonian}) is depicted in Fig. \ref{fig:chain}(a).
We will mostly focus on chains with n.n. and n.n.n couplings in this paper; 
but some results of chains with couplings beyond n.n.n. will also be presented.

We use the real-space renormalization-group method to study the 
Hamiltonian (\ref{hamiltonian}). The application of this method to
AF spin-1/2 chain with n.n. couplings only is well known. The
basic idea is to isolate the strongest bond in the system, decimate it, and
calculate the effective interactions generated between what were the 
third-nearest neighbors. The key simplifying features in this case are 
that the generated interactions
are always antiferromagnetic, and they connect only nearest-neighbor spins
(after the two spins coupled by the strongest bond are removed).

\begin{figure*}
\includegraphics[scale=0.6]{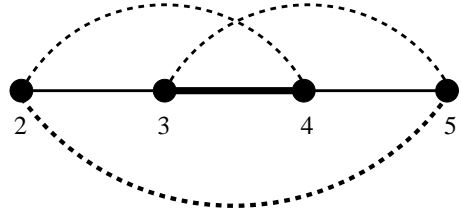}
\caption{The most complicated structure of a four-spin cluster where a given
spin is coupled to the other three spins.}
\label{fig:cluster}
\end{figure*}

Appropriate extensions of the original RG scheme need to be included in order
to study the present model with further neighbor couplings properly. 
First we notice that the coordination number, 
i.e. the number of spins coupled to a given spin, grows as the energy scale
is lowered so we need to keep track of the structure of the system. This is
in contrast to the AF spin-1/2 chain with n.n. couplings where the coordination
number is always 2. 
Second, as we will see later,
effective ferromagnetic couplings may be generated at certain stage as RSRG 
is carried out in the presence of antiferromagnetic n.n.n. couplings. 
The formation of
ferromagnetic couplings allows the possibility of generating effective spins
with sizes larger than 1/2, so we need to extend the RG rules 
to incorporate arbitrary spin sizes and coupling signs.  
Let us discuss these in more detail. Consider spin 3 and 4 in 
Fig. \ref{fig:chain}(a) which
are coupled by the strongest bond, and other spins in the system that couple to
at least one of them. Due to the presence of n.n.n. couplings, 
we have a six-spin problem instead of a four-spin problem for a given pair of 
spins coupled by the strongest bond. 
The Hamiltonian for the six-spin problem is given by :
\begin{equation}
H=H_0+H_I,
\end{equation}
where
\begin{eqnarray}
H_0 &=& J_{34} \boldsymbol{S}_3 \cdot \boldsymbol{S}_4, \nonumber \\
H_I &=& J_{23} \boldsymbol{S}_2 \cdot \boldsymbol{S}_3 + 
J_{45} \boldsymbol{S}_4 \cdot \boldsymbol{S}_5 + 
J_{13} \boldsymbol{S}_1 \cdot \boldsymbol{S}_3 \nonumber \\
&+& J_{35} \boldsymbol{S}_3 \cdot \boldsymbol{S}_5 +
J_{24} \boldsymbol{S}_2 \cdot \boldsymbol{S}_4 +
J_{46} \boldsymbol{S}_4 \cdot \boldsymbol{S}_4,
\end{eqnarray}
where $J_{ij}$ is the antiferromagnetic coupling between $\boldsymbol{S}_i$ 
and $\boldsymbol{S}_j$. We have shown in our previous work on the spin ladders
\cite{ek}
that to the second order perturbation calculation, $H_I$ only generates 
pairwise interactions among the spins and hence it is only necessary to 
include a pair of spins coupled to the two spins connected by the strongest 
bond when we consider the effective interaction between them, i.e. we just
have to consider four-spin clusters for a given segment which contains the 
strongest bond. Let us consider the most complicated four-spin cluster where
a given spin is coupled to three other spins as depicted in 
Fig. \ref{fig:cluster}. The renormalized coupling between two spins in the
cluster, say spins 2 and 5, is given by :
\begin{eqnarray}
\tilde{J}_{25} &=&  J_{25} + \frac{1}{2J_{34}} (J_{23}J_{45} + J_{24}J_{35} 
- J_{23}J_{35} - J_{24}J_{45}) \nonumber\\
&=& J_{25} + \frac{1}{2J_{34}} (J_{23} - J_{24})(J_{45} - J_{35})
\label{renorm}
\end{eqnarray}
where $\tilde{J}_{ij}$ is the {\em renormalized} coupling between 
$\boldsymbol{S}_i$ and $\boldsymbol{S}_j$, and $J_{ij}$ is the {\em original} 
bond between $\boldsymbol{S}_i$ and $\boldsymbol{S}_j$.
Examining Eq. \ref{renorm}, we can see that some of the contributions to the
renormalized coupling from 2nd order processes are {\em ferromagnetic}. 
The overall sign of the total interaction between the second and fifth
spin will be determined by the relative strength between the antiferromagnetic 
nearest-neighbor and next-nearest-neighbor bonds. 
In general if the n.n.n. couplings are very weak compared to 
the n.n. couplings then the ferromagnetic interactions will not appear. 
This is quite different from what we found in the 
study of the ladder where effective ferromagnetic interactions appear as soon
as the RG is applied to the system. Due to the possibility of the appearance 
of 
ferromagnetic couplings at some step of RG, it is necessary to generalize the 
RG procedure to include arbitrary spin sizes and coupling signs. The discussion
on how this is done has been spelled out in great detail in our earlier work
on the spin ladder. \cite{ek} We carry out the numerical calculation using 
the rules described in previous paragraphs and present the results in the next
section.

\section{Numerical Results}
We present numerical results for spin chains with nearest-neighbor (n.n.) and
next-nearest-neighbor (n.n.n.) interactions with total number of spins up to 
60000. We search for the bond with the largest gap, $\Delta_0$, which is 
defined
as the gap between the ground state and the first excited state, decimate it, 
and calculate the effective interactions among the remaining spins. The 
procedure is repeated until the number of spins left is about 1\% of the 
original number of spins in the system. We use 100 samples and take the 
disorder average over all these samples in all our calculations. 
The nearest-neighbor bonds are chosen 
to be distributed randomly according to the power-law probability distribution:
\begin{equation}
\label{distribution}
P_{n.n.}(J_{i}) = (1-\alpha) J_{i}^{-\alpha}, 0 < J_{i} < 1,
\end{equation}
where the power-law exponent $\alpha < 1$ parametrizes the randomness strength;
the larger $\alpha$, the stronger the randomness.
The reason for choosing a power-law form is because for the random spin-1/2
chain, the fixed point distribution is known to be in the power-law form. So
by choosing initial distributions in the power-law form, we expect to start
closer to the fixed point and hence reduce the necessity to use larger system
size. 

We consider two different ways
of generating the n.n.n. bonds. First we consider n.n.n. bonds which
are completely
correlated with the n.n. bonds, where next-nearest-neighbor bond $K_i$ is
determined from the n.n. bonds through the following relation:
\begin{equation}
K_i = \Lambda \frac{J_i J_{i+1}}{\Omega_0},
\label{nnn}
\end{equation}
where $\Lambda$ is a parameter introduced to  
control the strength of
next-nearest-neighbor interactions and $\Omega_0$ is the cutoff
of the initial nearest-neighbor bonds distribution, which is 1. In the limit 
$\Lambda \to 0$, the AF spin-1/2 chain with nearest-neighbor interactions only 
is recovered. 
Eq. (\ref{nnn}) comes from the following consideration.
The interactions between two spins come from
the overlap integral of the electron wave functions which are bound to the 
atoms sitting
on the lattice sites. In general, the wave function decays exponentially at
large distances, and so does the overlap integral. 
Let us consider three electrons sitting on different lattice 
sites labeled 1,2, and 3. For two electrons separated by
a distance $R$, the typical interaction would have a form of $J \sim e^{-R/a}$,
where $a$ is a length scale of order the size of the wave function.
Based on this picture, the interaction between the first and third spins, which
is basically the overlap integral between the first and third spins, can be
written as $J \sim e^{-(R_3 - R_1)/a}$, where $R_3$ and $R_1$ are measured 
with 
respect to some reference point. This relation can be rewritten as :
\begin{equation}
J \sim e^{-(R_3 - R_2)/a} e^{-(R_2 - R_1)/a} \propto J_2 J_1,
\end{equation}
where $J_i$ is the overlap integral between $S_i$ and $S_{i+1}$. 
Hence, it is reasonable to model the correlation as the product of two 
nearest-neighbor bonds as shown in Eq. (\ref{nnn}). We focus mostly on this
type of further neighbor coupling, and unless stated otherwise, the results
presented below are for this type of further neighbor coupling. 
For comparison, we have also studied cases in which the n.n.n. couplings are
uncorrelated case with the n.n. couplings, i.e. 
the n.n.n. bonds are distributed randomly in the
system, independent of of the distribution of the n.n. bonds. We choose the 
distribution to be in a power-law form with the same exponent, but a 
different cutoff $\Lambda$ :
\begin{equation}
P_{n.n.n.}(K_i) = \frac{1-\alpha}{\Lambda^{1-\alpha}} K_{i}^{-\alpha},
0 < K_i < \Lambda.
\end{equation}
Again $\Lambda$ parametrizes the strength of n.n.n. couplings. As we will see 
later, while the topology of the phase diagrams are the same for these two
cases, there is huge quantitative differences in the position of the phase
boundary.

\begin{figure*}
\includegraphics[scale=0.7,angle=-90]{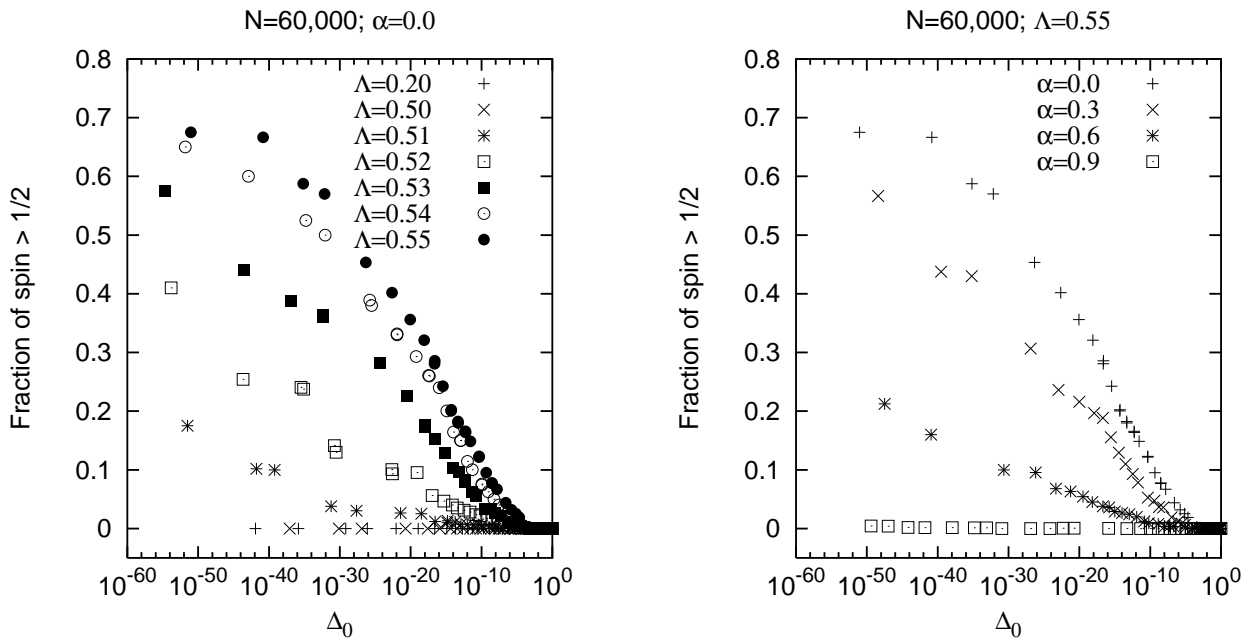}
\caption{The sample-averaged fraction of spins larger than 1/2 as a function 
of energy scale, $\Delta_0$. The error bars are about the size of the data 
points shown in the figure.
The left panel shows how the fraction of spins larger than 1/2 
for $\alpha=0.0$ changes as $\Lambda$ is varied and the right panel for 
$\Lambda=0.55$ as $\alpha$ is varied. Both are calculated for N=60,000. Strong
enough next-nearest-neighbor interactions will drive the system into a new
phase controlled by large effective spins. All calculations are done with 
correlated next-nearest-neighbor bonds given in Eq. (\ref{nnn}).}
\label{fig:spin}
\end{figure*}

As we carry out RSRG numerically, we monitor the appearance and
proliferation of large effective spins in the system.
We plot the sample-averaged fraction of spins larger 
than 1/2 as a function of energy scale, $\Delta_0$, in Fig. \ref{fig:spin}.
The left panel of Fig. \ref{fig:spin} shows how the formation of large 
effective spins evolves as the energy scale, $\Delta_0$, is lowered
by fixing $\alpha=0$ and varying n.n.n. bond strength 
controlled by 
$\Lambda$, while the right one by fixing $\Lambda=0.55$ and varying $\alpha$.
Let us analyze the left panel of Fig. \ref{fig:spin}. It is very clear that, 
for fixed $\alpha$, different antiferromagnetic n.n.n. bond strength will lead 
to different scenarios in the low energy limit. 
For weak enough $\Lambda$ (in the regime where $\Lambda < 0.5$) 
we do not find spin sizes other
than 1/2; not only we never find any spin larger than 1/2 but also we never
find any ferromagnetic bonds in this regime. The situation drastically changes
when we tune the strength of antiferromagnetic n.n.n. bonds up to 0.55 where 
we can see clearly
that large effective spins dominate in the low energy limit and drive the 
system into a new phase. This can be 
understood in the following way. For weak enough n.n.n. bonds, these 
interactions are always suppressed by the presence of n.n. bonds. We have 
explained in Eq. (\ref{renorm}) that the ferromagnetic bond will 
appear if the n.n.n. bonds are strong enough to overcome the n.n. bonds. 
Apparently for $\Lambda < 0.5$, the n.n.n. bonds are too weak to compete with
n.n. bonds so we never see the emergence of ferromagnetic interactions in the
system. On the other hand, for $\Lambda > 0.55$, the antiferromagnetic n.n.n.
bonds are strong enough to overcome the n.n. bonds and allow the 
appearance of ferromagnetic bonds which in turn will drive the system into a 
new phase controlled by large effective spins.

The right panel of Fig. \ref{fig:spin} shows another study of
how large effective
spins appear in the system by varying the disorder strength $\alpha$ for fixed
$\Lambda = 0.55$. We find that the formation of large effective spins is 
suppressed as the bond disorder gets stronger. This also has a simple 
explanation. With increasing bond disorder strength, the probability of finding
weak n.n. bonds is getting bigger. This will give us even weaker n.n.n. bonds
because of the correlation between a next-nearest-neighbor bond with two
nearest-neighbor bonds as given by Eq. \ref{nnn}. These weak n.n.n. bonds
can not compete with the n.n. bonds which in turn will suppress the formation
of ferromagnetic bonds in the system. Based on this view, we can 
understand why large effective spins are more difficult to form in the 
regime where the bond disorder is strong. So for strong enough
bond disorder, no ferromagnetic bonds will appear due to the fact that n.n.n.
bonds could not compete with n.n. bonds and the system will remain in the
Random Singlet (RS) phase.

\begin{figure*}
\includegraphics[scale=0.7,angle=-90]{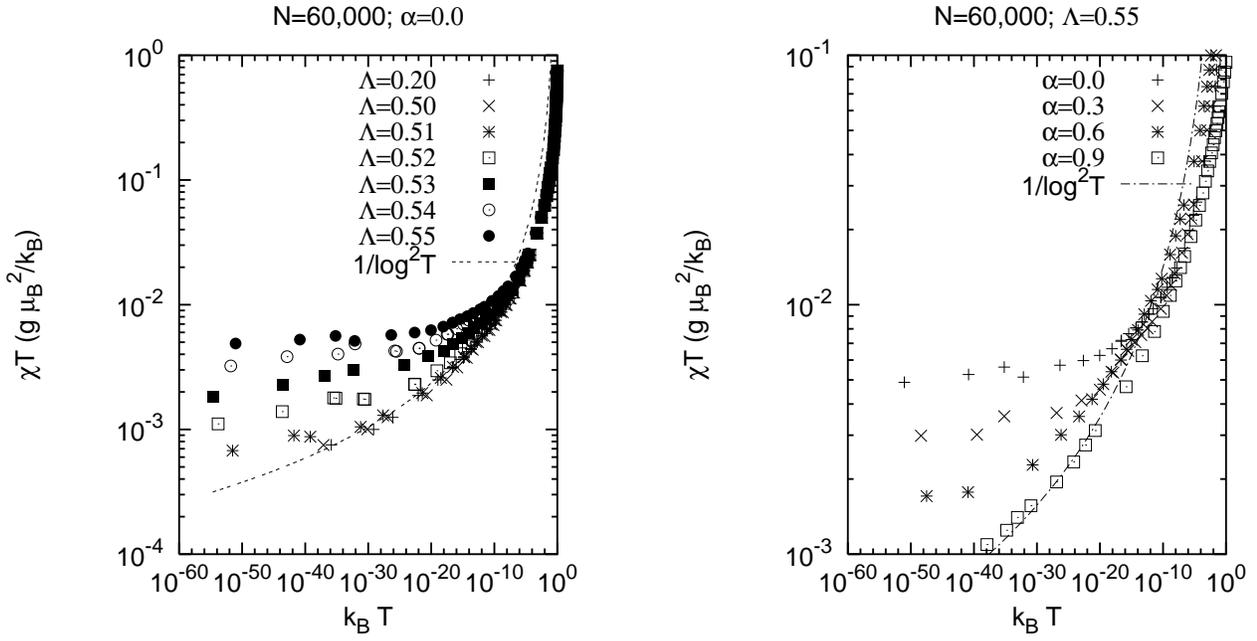}
\caption{The sample-averaged $\chi T$ as a function of parameters 
of the model, $\alpha$ and $\Lambda$.  The error bars are about the size of 
the data points. For 
strong enough correlated next-nearest-neighbor interactions, given in 
Eq. (\ref{nnn}), the susceptibilities behave
differently from $1/T log^2 T$. The contribution to the susceptibilities come
from large effective spins formed at low temperature.}
\label{fig:sus}
\end{figure*}

The appearance of a new phase can also be deduced from plotting the 
sample-averaged $\chi T$ as a function of temperature where the temperature
is associated with the energy scale, $\Delta_0$. 
We plot this in Fig. \ref{fig:sus} where in the left panel
$\alpha$ is fixed and $\Lambda$ is varied, whereas in the right panel 
$\Lambda$ is fixed and $\alpha$ is varied. $\chi T$ in the
RS phase is well known to be given by $1/log^2T$.
For fixed $\alpha=0$, we can see increasing deviations from $1/log^2 T$ 
with increasing strength for $\Lambda$ which gives us a clear indication that 
the system is driven away from the RS phase; for $\lambda > 0.5$ instead of 
falling as $1/log^2T$, $\chi T$ appears to approach a constant in the low-$T$
limit. The explanation for this 
behavior is similar to the discussion in the previous paragraph. Strong enough
n.n.n. bonds will allow the appearance of ferromagnetic bonds which in turn
will form large effective spins in the low energy limit. These strongly 
correlated effective spins govern the susceptibility of the system at low 
temperature. The susceptibility in this phase has different origin from
the susceptibility for the RS phase where the contribution comes from the 
undecimated half spins. The same situation is encountered when $\Lambda$ is 
fixed and $\alpha$ is varied as shown on the right panel of 
Fig. \ref{fig:sus}. The deviations are more significant for small $\alpha$.
This is consistent with our discussion on the previous paragraph that for
strong enough bond disorder, the system will remain in the RS phase because
the overall strength of n.n.n. bonds is much weaker than that of n.n. bonds. 
This is indeed what we see in our numerical results that $\chi T$ for 
bigger $\alpha (> 0.6)$ is closer to the value for RS phase $1/log^2T$.

\begin{figure*}
\includegraphics[scale=0.7,angle=-90]{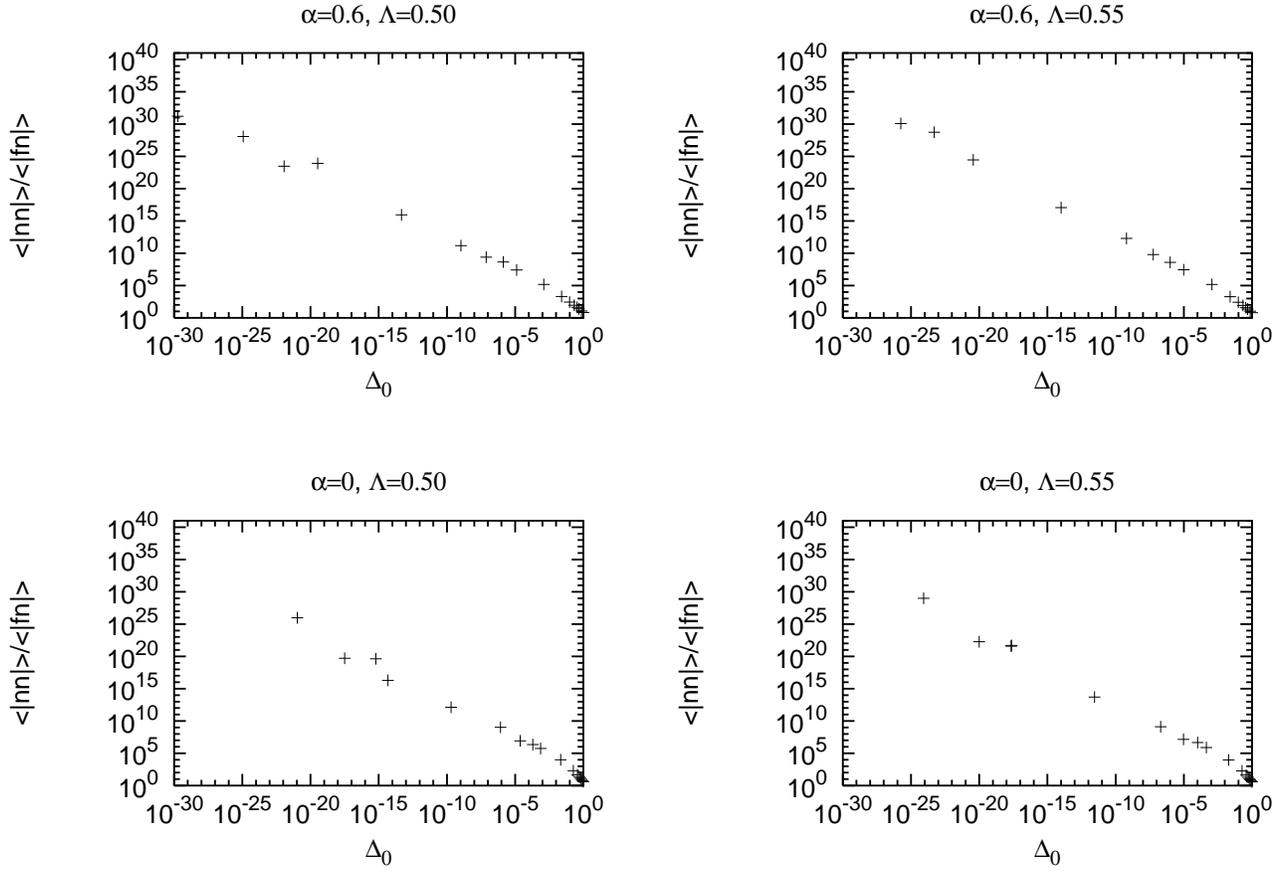}
\caption{The sample-averaged ratio of the strength of the nearest-neighbor 
bonds to the 
strength of the bonds that are beyond nearest neighbor as a function of energy
scale. It is clear from the plot that in either side of the phase, the 
interactions are dominated by nearest-neighbor bonds only. We use the 
correlated next-nearest-neighbor interactions defined in Eq. (\ref{nnn}).}
\label{fig:ratio}
\end{figure*}

\begin{figure*}
\includegraphics[scale=0.75,angle=-90]{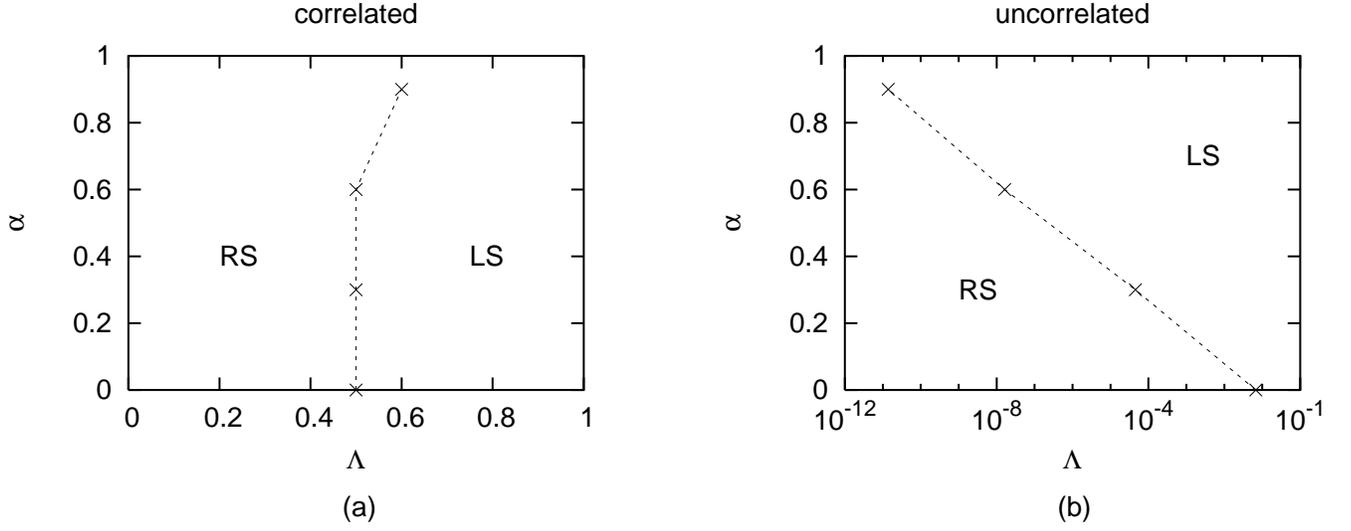}
\caption{(a) The numerically determined phase diagram for spin chains with 
competing 
interactions between nearest-neighbor and next-nearest-neighbor interactions.
The n.n.n. interactions are correlated with the n.n. interactions (see text). 
(b) The numerically determined phase diagram for spin chains with uncorrelated 
n.n.n. interactions. 
In both
cases $\alpha$ denotes the strength of the bond randomness and $\Lambda$ 
represents the strength of the next-nearest-neighbor interactions. The crosses
in both figures are obtained from numerical calculations. The dashed lines are
drawn by connecting the data points to see the phase boundary more clearly.}
\label{fig:phase}
\end{figure*}

We have established that there two phases in the system. The transition from
one phase to another is 
controlled by the strength of bond disorder $\alpha$ and the strength of n.n.n.
bonds $\Lambda$. For $\alpha=0$ and $\Lambda < 0.5$ the system remains in the 
RS phase while for $\Lambda > 0.55$ the system is driven into the new phase.
We have already seen that the new
phase is controlled by large effective spins in the low energy limit. Is there
any other parameter we can use to study the nature of of the new phase?
We address this question by studying the ratio of n.n. bond strength to 
n.n.n. bond strength in the two phases as shown in Fig. \ref{fig:ratio}. It is
found that on either side of the phase boundary, nearest-neighbor bonds always 
dominate further-neighbor bonds.\cite{further} Now we have a more 
complete picture of the new phase found in the system. The new phase is 
controlled by large effective spins in the low energy limit and the dominant
interactions come from the nearest-neighbor bonds only. These nearest-neighbor
interactions consist of both antiferromagnetic and ferromagnetic bonds. These
results suggest that in the low energy limit, spin chains with 
antiferromagnetic n.n. and sufficiently strong n.n.n. interactions behave 
just like
random antiferromagnetic-ferromagnetic spin chains, including a Curie 
susceptibility discussed earlier. This brings us to the 
conclusion that the new
phase found in the system we are studying is the same as the large spin
phase found in
the random antiferromagnetic-ferromagnetic spin chains in the low energy limit.
The numerically determined phase diagram for spin chains
with random antiferromagnetic n.n. and n.n.n. bonds is shown in 
Fig. \ref{fig:phase}.


The left panel of this figure shows the phase diagram for the correlated
next-nearest-neighbor bonds as given by Eq. (\ref{nnn}) whereas the right
panel for uncorrelated next-nearest-neighbor bonds. In both cases we 
find that the system supports only two phases, which are the Random Singlet
phase and the Large Spin phase. There are some differences of the phase
boundaries on these two cases. First, the trend on how the phase boundaries 
change as we vary $\alpha$ and $\Lambda$ is different for the correlated and
uncorrelated next-nearest-neighbor bonds. For the correlated case, $\Lambda$ 
stays constant as we increase $\alpha$ from 0 to 0.6 and tends to increase
for $\alpha$ larger than 0.6. For the uncorrelated one, $\Lambda$ decreases
with increasing $\alpha$. Secondly, 
the magnitude of critical $\Lambda$ for the uncorrelated n.n.n. couplings 
is much smaller than that for the correlated case, by as much as 10 orders
of magnitude for $\alpha$ close to 1!

We believe the these differences can be understood as the following. 
For the uncorrelated case we assign a
probability distribution function
for the n.n.n. bonds whose cutoff is determined by 
$\Lambda$, and the bonds are generated independent of the configuration
of the n.n. bonds. Although in 
general the strength of the n.n.n. bonds is much weaker than that of n.n. 
bonds when $\Lambda$ is small, 
due to the uncorrelated nature of the way they are generated, there is a 
small probability that the next-nearest-neighbor coupling 
is actually stronger than the nearest-neighbor ones in some region of the
system. 
As we have explained earlier in
the text, the overall sign of the total interaction generated by RG
between two spins depends
heavily on the relative strength of the antiferromagnetic n.n. and n.n.n. 
bonds; thus such rare events can lead to the generation of ferromagnetic bonds,
which in turn may proliferate as energy scale goes down. In the correlated 
case, on the other hand, such rare events are greatly suppressed by the 
correlation between n.n. and n.n.n. bonds. 
We also know that $\alpha$ parametrizes the width of the distribution;
for a given $\Lambda$ in the uncorrelated case, the bigger $\alpha$ is, the
wider the distributions are for both the n.n. and n.n.n. bond distributions, 
thus
the larger the probability of the rare events discussed above are, and the
more likely ferromagnetic couplings get generated. On the other
hand this effect is again suppressed for the case of correlated n.n.n. bonds,
due to the way we parametrizes their strength; the larger $\alpha$ is, the 
smaller the overall strength of the n.n.n. bonds due to the way the are 
generated.

\begin{figure*}
\includegraphics[scale=0.75,angle=-90]{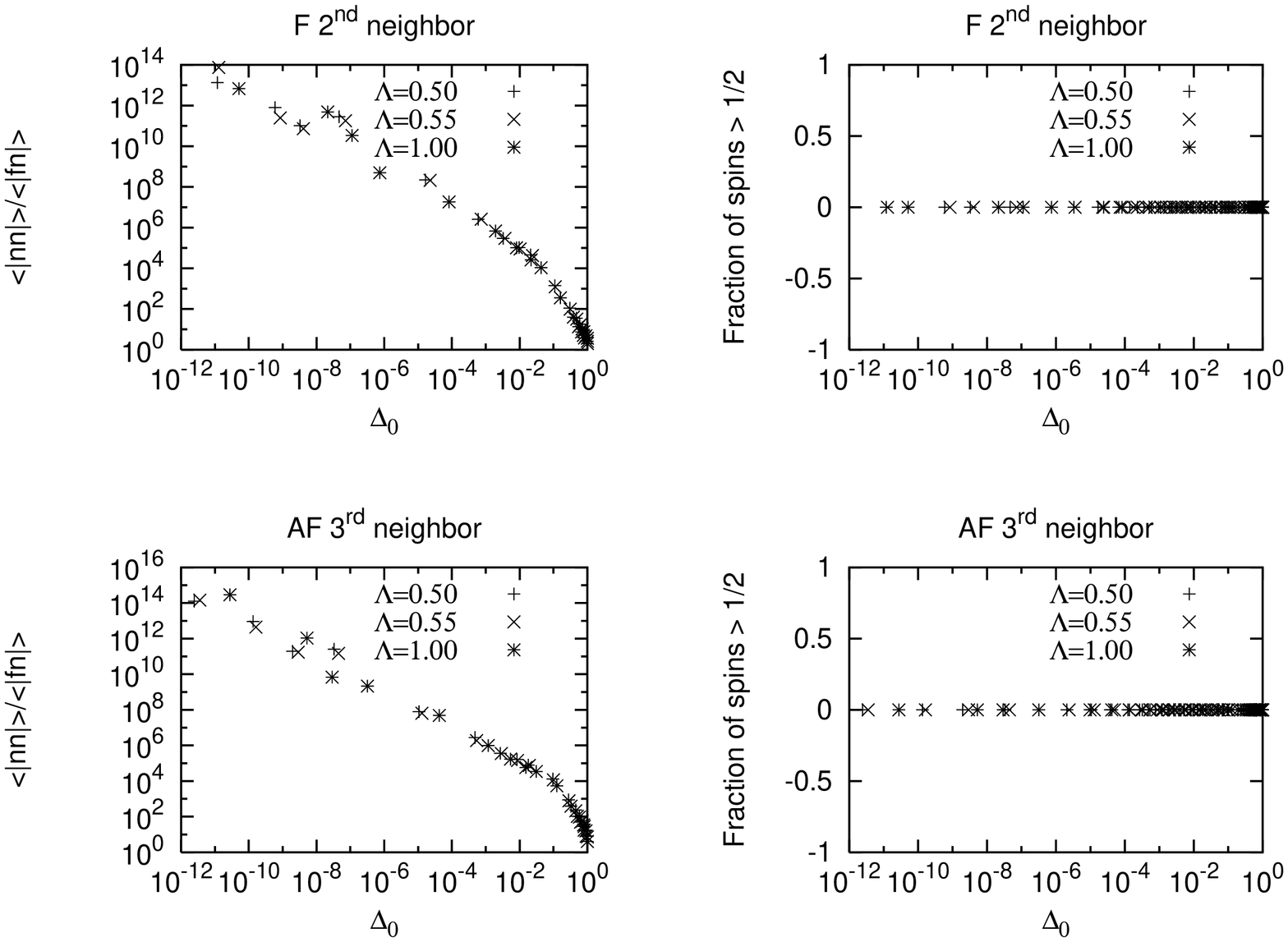}
\caption{The sample-averaged ratio of the strength of the nearest-neighbor 
bonds to the 
strength of the bonds that are beyond nearest neighbor and the fraction of 
spins larger than 1/2 as a function of energy scale as a function of energy
scale for the model with no frustration introduced into the system.
Two types of interactions
which do not generate frustration, e.g. ferromagnetic second 
neighbors and antiferromagnetic third neighbors, are introduced into the 
system. The upper two panels show the calculation for a model in which 
ferromagnetic second-neighbor interactions are introduced into the system while
the lower two panels for antiferromagnetic third-neighbor interactions. All
graphs are calculated for $\alpha=0$ but with varying $\Lambda$. It is clear 
from the plot that the interactions are dominated by nearest-neighbor bonds 
only regardless the value of $\Lambda$ and there is no formation of effective 
spins whose sizes are larger than 1/2.}
\label{fig:nofrus}
\end{figure*}

As discussed earlier, the appearance of effective ferromagnetic couplings is a
consequence of competition between nearest and further neighbor couplings, 
or frustration.
We have also studied spin chains with further-neighbor interactions
which do {\em not} introduce frustration to the system. This can be done by 
introducing ferromagnetic next-nearest-neighbor bonds or antiferromagnetic
third-nearest-neighbor bonds. The ferromagnetic next-nearest-neighbor bonds 
and antiferromagnetic third-nearest-neighbor bonds are generated in the same
way as discussed at the beginning of this section, i.e. the bonds are generated
through Eq. (\ref{nnn}).
We present our results for this particular
system in Fig. \ref{fig:nofrus}. 

The upper panels of Fig. \ref{fig:nofrus} 
show the sample-averaged plot of the strength of nearest-neighbor interactions 
compared to
the strength of further-neighbor interactions and the fraction of spins with 
sizes larger than 1/2 as 
a function of energy scale, $\Delta_0$, for the system with ferromagnetic
next-nearest-neighbor bonds. We choose to fix $\alpha=0$ and to vary 
$\Lambda$ to see how the ratio changes as the energy scale is lowered. We find
that the nearest-neighbor interactions always dominate over 
further-neighbor interactions at all energy scale. 
The evolution of the spin sizes as the energy scale is lowered is also studied
here. The result shows that no spin having sizes larger than 1/2 is found in
the system. Based on these results we conclude that the presence of 
ferromagnetic next-nearest-neighbor bonds does not drive the system into a new
phase. The couplings are dominated by antiferromagnetic bonds which suppress
the formation of effective spins larger than 1/2 at low energy. In the low 
energy limit the system stays in the RS phase.
The lower panels of Fig. \ref{fig:nofrus} 
show the plot of the ratio between the strength of nearest and further neighbor
bonds and the fraction of spins with sizes larger than 1/2 as 
a function of energy scale, $\Delta_0$, for the system with antiferromagnetic
third-nearest-neighbor bonds. We also fixed $\alpha=0$ and vary $\Lambda$ for 
this case. The results are the same for those with 
ferromagnetic next-nearest-neighbor bonds. These results give us a strong 
indication that the system stays in the RS phase. We can thus conclude that 
non-frustrating further neighbor bonds act as irrelevant perturbations in the 
low energy 
limit, and hence the system stays in the RS phase.

\section{Summary and discussion}

In this paper we have used the real space renormalization group method to study
random antiferromagnetic spin-1/2 chains, with both nearest- and further
neighbor interactions. We find that the system supports two phases, the 
random singlet phase and the large spin phase. The latter is only stabilized
by sufficiently strong further neighbor couplings that compete with the nearest
neighbor couplings, so that there is frustration in the system.

The real space renormalization group  procedure
is quantitatively accurate only when the 
initial distributions of the couplings are broad. We believe, however, our 
conclusion remains valid even if the initial distribution of couplings is 
not broad. In the case of nearest neighbor coupling only, 
Doty and Fisher\cite{doty} showed that weak bond randomness is a relevant 
perturbation that immediately destabilizes the Luttinger liquid
fixed point that describes the gapless phase of the pure chain, and bond 
randomness {\em grows} as energy scale goes down, eventually brings the 
system to the random singlet fixed point. Their arguments remain valid even
in the presence of further neighbor couplings, as long as they are not strong
enough to destabilize the gapless phase in the absence of bond randomness.
On the other hand when they are strong enough to put the pure system in the
gapped phase with spontaneous dimerization, one of us\cite{yhbg} showed that 
the dimerized phase is {\em also unstable} against weak randomness, as 
randomness nucleates solitons and destroys spontaneous dimerization; the 
low-energy degrees of freedom are the half-spins carried by the solitons, with
random interaction with broad distribution (due to the fluctuation of inter
soliton distance etc). Depending on whether the coupling between these spins
are purely AF or both F and AF, the systems can be in either one of the two 
phases we find here. We thus conclude these are the only two phases the system
supports in the presence of any amount of bond randomness.\cite{note} 
     
The frustration induced ferromagnetic coupling and the resultant large spin 
formation has been discussed in a different context before.\cite{yb}
In that work Yang and Bhatt studied 
spin-1 chains with random AF nearest neighbor 
bonds, with both quadratic and bi-quadratic couplings on each bond. 
It was shown that even though overall each individual bond is AF, as long as 
in some of the bonds 
the quadratic and bi-quadratic couplings have opposite tendency (i.e., one AF
and the other F), effective ferromagnetic couplings may be generated at 
low-energy, and the large spin phase stabilized. In this case the bonds are 
frustrated due to the competition between quadratic and bi-quadratic couplings
on the {\em same} bond. Thus the phenomena of 
frustration induced large spin formation, although never seen in pure systems,
may actually be rather generic in disordered systems.

\acknowledgements

We have benefited greatly from very useful discussions with R. N. Bhatt.
This work was supported by
NSF grant No. DMR-0225698, the Center for Materials Research and Technology 
(MARTECH), and the A. P. Sloan Foundation.

\begin{thebibliography}{99}
\bibitem{further} As RG is applied to the chain, the coordination number grows,
and a given spin may couple to many other spins so at this point we divide the
interactions into two types: nearest-neighbor bonds and those other than
nearest-neighbor bonds, i.e. further-neighbor bonds.
\end{thebibliography}
\end{document}